\documentclass[12pt]{iopart}

\usepackage{iopams,graphicx}
\begin{document}

\title{Replica analysis of Franz-Parisi potential for sparse systems}

\author{Masahiko Ueda}
\address{Department of Physics, Kyoto University, Kyoto 606-8502, Japan}
\ead{ueda@ton.scphys.kyoto-u.ac.jp}
\author{Yoshiyuki Kabashima}
\address{Department of Computational Intelligence and Systems Science, Tokyo Institute of Technology, Yokohama 226-8502, Japan}
\ead{kaba@dis.titech.ac.jp}

\begin{abstract}
We propose a method for calculating the Franz-Parisi potential for spin glass models on sparse random graphs using the replica method under the replica symmetric ansatz.
The resulting self-consistent equations have the solution with the characteristic structure of multi-body overlaps, and the self-consistent equations under this solution are equivalent to the one-step replica symmetry breaking (1RSB) cavity equation with Parisi parameter $x=1$.
This method is useful for the evaluation of transition temperatures of the $p$-spin model on regular random graphs under a uniform magnetic field.
\end{abstract}


\maketitle

\section{Introduction}
\label{sec:Intro}
The Franz-Parisi potential (FPP) \cite{FP1995, FP1997} is defined as an effective potential of overlap $q$ between two replicas in two temperatures, $T$ and $T^\prime$.
This concept was originally introduced for fully connected spin glass models in order to characterize the one-step replica symmetry breaking (1RSB) as an appearance of the metastable states of a thermodynamic potential.
A primary advantage of the FPP framework is the ability to describe the 1RSB under the replica symmetric (RS) calculation's level.
In addition to its technical ease, this method offers a useful physical insight into what occurs when the replica symmetry is broken.
The FPP can also be used to characterize the temperature chaos \cite{BFP1997}, to determine the phase diagram of finite-dimensional spin glass \cite{CPPS1990} and structural glass \cite{FP1998}, and to detect instability in a ground state in response to perturbations of a certain type \cite{PPR2003}.

The extension of the FPP framework to sparsely connected systems continued over the next decade.
In \cite{ZdeKrz2010}, a methodology for evaluating the FPP of sparsely connected systems was developed using the cavity method and reinterpreted as the states-following method.
The FPP's role may be more significant in sparsely connected systems than in fully connected systems.
In general, we need to solve functional equations for ``distributions of distributions'' in the standard 1RSB framework of the sparse systems \cite{MezMon}.
However, the equivalence between the FPP framework and the 1RSB framework with Parisi parameter $x=1$ \cite{MezMon2006, MRS2008, Zde2009} allows us to evaluate the various quantities simply by solving self-consistent equations for ``distributions'' utilizing population dynamics.
This enables us to accurately determine the conditions of dynamical and static transitions from the RS solution in a computationally efficient manner.
Because no other methods outperform the accuracy of this evaluation given the same computational resources, the FPP is an indispensable component of the analysis of the glassy behavior of sparse systems.

A primary objective of this paper is to derive the FPP of sparse systems using the replica method \cite{MezMon}.
One of our motivations is to fill the technical gap between the derivations for the fully and sparsely connected systems.
Addressing both systems in a unified manner will help deepen our understanding of the FPP technique.
Our other motivation is to provide an FPP with another basis.
As the cavity method significantly relies on the tree-like nature of random graphs, incorporating the loops' effects is difficult when the objective system is defined on a loopy graph, such as a finite-dimensional lattice.
Providing another basis using the replica method may help us address systems on loopy graphs, although this paper's focus will remain on random graphs.
In addition, we provide a general and explicit proof for the equivalence between the FPP and 1RSB with $x=1$ for sparse systems in the presence of external fields, showing nontrivial identities of multi-spin overlaps that reflect the ultrametric structure.

This paper is organized as follows.
In section \ref{sec:model}, we establish models and introduce the Franz-Parisi potential and its Legendre transform.
In section \ref{sec:FP}, we develop a method to calculate the Legendre transform of the FPP using the replica method under the replica symmetric ansatz.
In section \ref{sec:1rsb}, we demonstrate that the calculation of the Legendre transform of the Franz-Parisi potential is equivalent to the 1RSB cavity method with $x=1$ for $T^\prime=T$ and $h_\mathrm{ext}=0$ by introducing an ultrametric structure.
In section \ref{sec:numerical}, we report the numerical results obtained by applying this method to the calculation of the 1RSB transition temperatures.
The final section summarizes our study.

\section{Model setup}
\label{sec:model}
We consider the $p$-spin model of size $N$ on regular random graphs under a uniform magnetic field.
The spin variable $\sigma_i \in \left\{ -1, 1 \right\}$ is defined for each vertex $i$, and we collectively write the set of these variables as $\textrm{\boldmath $\sigma$} \equiv \left\{ \sigma_i \right\}_{i=1}^N$.
The Hamiltonian is given by
\begin{eqnarray}
 H_0 (\textrm{\boldmath $\sigma$}) &=& - \sum_{i_1< \cdots< i_p} g_{i_1, \cdots, i_p} J_{i_1, \cdots, i_p} \sigma_{i_1}\cdots \sigma_{i_p} - h_0 \sum_{i=1}^N \sigma_i.
\end{eqnarray}
Here, $g_{i_1, \cdots, i_p}$ is a random variable that represents the lack or existence of a connection with the value $0$ or $1$, respectively, 
$J_{i_1, \cdots, i_p}$ is a random interaction variable, and
$h_0$ is a uniform magnetic field.
In particular, we consider the $\pm J$-type $p$-spin model on a $C$-regular random graph.
The probability distribution of $g_{i_1, \cdots, i_p}$ for this model is given by
\begin{eqnarray}
 P\left( \left\{ g_{i_1, \cdots, i_p} \right\} \right) &=& \prod_{i=1}^N \delta \left( \sum_{i_1< \cdots< i_{p-1}}g_{i_1, \cdots, i_{p-1}, i} - C \right).
 \label{eq:regular}
\end{eqnarray}
In addition, the probability distribution of $J_{i_1, \cdots, i_p}$ is
\begin{eqnarray}
 P(J) &=& \frac{1}{2} \left[ \delta_{J, 1} + \delta_{J, -1} \right].
\end{eqnarray}
Let us consider the case in which $g_{i_1, \cdots, i_p}$ and $J_{i_1, \cdots, i_p}$ are symmetric under index permutations.
The $h_0=0$ case has been frequently analyzed \cite{FLRZ2001, FMRWZ2001, MonRic2004, NakHuk2009}.
Specifically, it has been shown that the low temperature phase is reasonably described by the 1RSB solution.
However, the $p$-spin model on sparse random graphs under a nonzero uniform magnetic field has not been examined.

Now we introduce an effective potential of overlap: the Franz-Parisi potential.
First, we consider a free energy of the system $\textrm{\boldmath $\sigma$}$ in the situation that its overlap with the spin configuration $\textrm{\boldmath $s$}$ equals $q$.
This free energy is expressed as
\begin{eqnarray}
 -\frac{1}{\beta} \log\left( \sum_{\textrm{\boldmath $\sigma$}} e^{-\beta H_0(\textrm{\boldmath $\sigma$})} \delta\left( q-\frac{1}{N}\sum_{i=1}^N \sigma_is_i \right) \right).
\end{eqnarray}
We assume that $\textrm{\boldmath $s$}$ obeys the canonical distribution with Hamiltonian $H_0 (\textrm{\boldmath $s$})$ and inverse temperature $\beta^\prime$.
The Franz-Parisi potential is defined as the configurational average of this restricted free energy with respect to spin configuration $\textrm{\boldmath $s$}$ and random variables $\left\{ g_{i_1, \cdots, i_p} \right\}$ and $\left\{ J_{i_1, \cdots, i_p} \right\}$:
\begin{eqnarray}
 \fl - \beta v(\beta, \beta^\prime, q) &\equiv& \lim_{N\rightarrow \infty} \frac{1}{N} \mathbb{E}_g \mathbb{E}_J \left[ \frac{1}{Z_0^\prime} \sum_{\textrm{\boldmath $s$}} e^{-\beta^\prime H_0(\textrm{\boldmath $s$})} \right. \nonumber \\
 \fl && \qquad \left. \times \log\left( \sum_{\textrm{\boldmath $\sigma$}} e^{-\beta H_0(\textrm{\boldmath $\sigma$})} \delta\left( q-\frac{1}{N}\sum_{i=1}^N \sigma_is_i \right) \right) \right],
 \label{eq:FP_potential}
\end{eqnarray}
where $Z_0^\prime \equiv \sum_{\textrm{\boldmath $s$}} e^{-\beta^\prime H_0(\textrm{\boldmath $s$})}$, and $\mathbb{E}_g$ and $\mathbb{E}_J$ represent the expected values with respect to $\left\{ g_{i_1, \cdots, i_p} \right\}$ and $\left\{ J_{i_1, \cdots, i_p} \right\}$, respectively.
Because it is sufficient to use the Legendre transform of this quantity in order to evaluate transition temperatures, we consider its Legendre transform.
(We briefly mention the Franz-Parisi potential itself in \ref{ap:potential}.)
Let us introduce a field $h_\mathrm{ext}$ that is conjugate to overlap $q$.
If a self-averaging property is justified, the Legendre transform of the Franz-Parisi potential is given by
\begin{eqnarray}
 \fl - \beta g(\beta, \beta^\prime, h_\mathrm{ext}) &\equiv& \lim_{N\rightarrow \infty} \frac{1}{N} \mathbb{E}_g \mathbb{E}_J \left[ \frac{1}{Z_0^\prime} \sum_{\textrm{\boldmath $s$}} e^{-\beta^\prime H_0(\textrm{\boldmath $s$})} \right. \nonumber \\
 \fl && \qquad \left. \times \log\left( \sum_{\textrm{\boldmath $\sigma$}} e^{-\beta H_0(\textrm{\boldmath $\sigma$}) + \beta h_\mathrm{ext} \sum_{i=1}^N \sigma_is_i} \right) \right].
 \label{eq:FP_free_energy}
\end{eqnarray}
In particular, we would like to analyze this quantity under the RS ansatz.

Due to a property of the Legendre transformation, considering the case $h_\mathrm{ext}=0$ corresponds to probing the minima of the Franz-Parisi potential.
In order to calculate the transition point, we need to set $\beta^\prime=\beta$. 
The transition point is the point where the local minima $g_\mathrm{SG}$ equals the global minima $g_\mathrm{para}$.
In other words, the transition temperature $T_\mathrm{K}$ is given by
\begin{eqnarray}
 -\beta g_\mathrm{SG}(\beta, \beta, 0) &=& -\beta g_\mathrm{para}(\beta, \beta, 0).
 \label{eq:FP_transition}
\end{eqnarray}

\section{Replica approach to Franz-Parisi potential}
\label{sec:FP}
\subsection{Evaluation for $n,m \in \mathbb{N}$}
Using the replica method, we can calculate the Legendre transform of the Franz-Parisi potential (\ref{eq:FP_free_energy}).
By introducing two replica numbers $m$ and $n$ as \cite{FP1995}, this quantity is expressed as
\begin{eqnarray}
 -\beta g(\beta, \beta^\prime, h_\mathrm{ext}) &=& \lim_{N\rightarrow \infty} \frac{1}{N} \lim_{m\rightarrow 0} \lim_{n\rightarrow 0} \frac{\partial }{\partial n} \mathbb{E} \left[ Z_{n,m} \right],
\end{eqnarray}
where the partition function of the $(n,m)$-replica system is defined to be
\begin{eqnarray}
 \fl \mathbb{E} \left[ Z_{n,m} \right] &\equiv& \mathbb{E}_g\mathbb{E}_J \left[ \sum_{\left\{ \textrm{\boldmath $s$}^{(a)} \right\}} \sum_{\left\{ \textrm{\boldmath $\sigma$}^{(b)} \right\}} e^{-\beta^\prime \sum_{a=1}^m H_0\left( \textrm{\boldmath $s$}^{(a)} \right) -\beta \sum_{b=1}^n H_0\left( \textrm{\boldmath $\sigma$}^{(b)} \right) + \beta h_\mathrm{ext} \sum_{b=1}^n \sum_{i=1}^N \sigma_i^{(b)} s_i^{(1)}} \right].
\end{eqnarray}
For simplicity, we introduce the notation $\vec{s}_i \equiv \left( s^{(1)}_i, \cdots, s^{(m)}_i \right)$ and $\vec{\sigma}_i \equiv \left( \sigma^{(1)}_i, \cdots, \sigma^{(n)}_i \right)$.
First, we define the quantity
\begin{eqnarray}
 N_G &\equiv& \sum_{\left\{ g_{i_1, \cdots, i_p} \right\}} \left\{ \prod_i \delta \left( \sum_{i_1< \cdots< i_{p-1}}g_{i_1, \cdots, i_{p-1}, i} - C \right) \right\},
 \label{eq:N_G}
\end{eqnarray}
which represents the number of bipartite graphs whose factor nodes and variable nodes uniformly have $p$ and $C$ links, respectively. 
By using a standard technique for analyzing sparsely connected systems, we obtain for large $N$
\begin{eqnarray}
 \fl \mathbb{E} \left[ Z_{n,m} \right] &\sim& \frac{1}{N_G} \left( \prod_{\vec{\sigma}, \vec{s}} N \int dm\left( \vec{\sigma}, \vec{s} \right) \int d\hat{m}\left( \vec{\sigma}, \vec{s} \right) \right) e^{-N\log C! -N \sum_{\vec{\sigma}, \vec{s}} \hat{m}\left( \vec{\sigma}, \vec{s} \right) m\left( \vec{\sigma}, \vec{s} \right)} \nonumber \\
 \fl && \qquad \times \left( \sum_{\vec{\sigma}, \vec{s}} \hat{m}\left( \vec{\sigma}, \vec{s} \right)^C e^{ \beta^\prime h_0 \sum_{a=1}^m s^{(a)} + \beta h_0 \sum_{b=1}^n \sigma^{(b)} + \beta h_\mathrm{ext} \sum_{b=1}^n \sigma^{(b)} s^{(1)}} \right)^N \nonumber \\
 \fl && \qquad \times \exp \left( \sum_{\vec{s}_{(1)}, \vec{\sigma}_{(1)}} \cdots \sum_{\vec{s}_{(p)}, \vec{\sigma}_{(p)}} \frac{N^p}{p!} 2^{m+n} \cosh^m\left( \beta^\prime \left| J \right| \right) \cosh^n\left( \beta \left| J \right| \right) \right. \nonumber \\
 \fl && \qquad \times \left\{ \prod_{j=1}^p m\left( \vec{\sigma}_{(j)}, \vec{s}_{(j)} \right) \right\} \nonumber \\
 \fl && \qquad \left. \times \mathbb{E}_J \left[ \prod_{a=1}^m \frac{1+\tanh\left( \beta^\prime J \right)\prod_{j=1}^p s_{(j)}^{(a)}}{2} \prod_{b=1}^n \frac{1+\tanh\left( \beta J \right)\prod_{j=1}^p \sigma_{(j)}^{(b)}}{2} \right] \right).
 \label{eq:znmcal}
\end{eqnarray}
See \ref{ap:znmcal} for details of the derivation.
A similar calculation can also be performed for $N_G$:
\begin{eqnarray}
 N_G &\sim& e^{N \left[ \frac{C}{p}\log\left( \frac{N^{p-1}C^{p-1}}{(p-1)!} \right) -\log C! - C + \frac{C}{p} \right]}.
 \label{eq:N_G_final}
\end{eqnarray}
(Details of the calculation is given in \ref{ap:N_G_cal}.)
Substituting this result into (\ref{eq:znmcal}) and evaluating the integral with respect to $m\left( \vec{\sigma},\vec{s} \right)$ and $\hat{m}\left( \vec{\sigma},\vec{s} \right)$ with saddle-point values $m_*\left( \vec{\sigma},\vec{s} \right)$ and $\hat{m}_*\left( \vec{\sigma},\vec{s} \right)$, we finally obtain
\begin{eqnarray}
 \mathbb{E} \left[ Z_{n,m} \right] &\sim& e^{-N\beta g_{n,m}}, 
\end{eqnarray}
with $g_{n,m}$ given by
\begin{eqnarray}
 \fl -\beta g_{n,m} &=& -\frac{C}{p} \log\frac{N^{p-1}C^{p-1}}{(p-1)!} + C - \frac{C}{p} - \sum_{\vec{\sigma},\vec{s}} m_*\left( \vec{\sigma},\vec{s} \right) \hat{m}_*\left( \vec{\sigma},\vec{s} \right) \nonumber \\
 \fl && + \log\left(\sum_{\vec{\sigma},\vec{s}} \hat{m}_*\left( \vec{\sigma},\vec{s} \right)^C e^{\beta h_\mathrm{ext} \sum_{b=1}^n\sigma^{(b)}s^{(1)} + \beta^\prime h_0 \sum_{a=1}^m s^{(a)} + \beta h_0 \sum_{b=1}^n\sigma^{(b)}} \right) \nonumber \\
 \fl && + \frac{N^{p-1}}{p!} 2^{m+n} \cosh^m\left( \beta^\prime J \right) \cosh^n\left( \beta J \right) \sum_{\vec{\sigma}_{(1)},\vec{s}_{(1)}} \cdots \sum_{\vec{\sigma}_{(p)},\vec{s}_{(p)}} \nonumber \\
 \fl && \qquad \times m_*\left( \vec{\sigma}_{(1)},\vec{s}_{(1)} \right) \cdots m_*\left( \vec{\sigma}_{(p)},\vec{s}_{(p)} \right) \nonumber \\
 \fl && \qquad \times \mathbb{E}_J \left[ \prod_{a=1}^m \frac{1+\tanh\left( \beta^\prime J \right)\prod_{j=1}^ps_{(j)}^{(a)}}{2} \prod_{b=1}^n \frac{1+\tanh\left( \beta J \right)\prod_{j=1}^p\sigma_{(j)}^{(b)}}{2} \right].
 \label{eq:gnmcal}
\end{eqnarray}
The functions $m_*\left( \vec{\sigma},\vec{s} \right)$ and $\hat{m}_*\left( \vec{\sigma},\vec{s} \right)$ correspond to order parameters for $(n,m)$-replicated sparse systems.

\subsection{Replica symmetric ansatz and analytic continuation to $n,m \in \mathbb{R}$}
The system $\vec{s}$ interacts with the system $\vec{\sigma}$ only through $s^{(1)}$.
Because spin reversal symmetry is broken when the magnetic field $h_0$ is present, it is natural to introduce cavity fields that are conditioned by the value of $s^{(1)}$ when we assume replica symmetry.
It also follows that we should require the replica symmetric solution of the $\vec{s}$ system to be reproduced when we first take the limit $n\rightarrow 0$. This means that $s^{(1)}$ is no longer special.
In consideration of these two requirements, we introduce the RS ansatz as
\begin{eqnarray}
 \fl m_*\left( \vec{\sigma}, \vec{s} \right) &=& \alpha \int d\rho \left(h \right) d\mu \left(w|h, s^{(1)} \right) \prod_{a=1}^m \frac{1+s^{(a)}\tanh(\beta^\prime h)}{2} \prod_{b=1}^n \frac{1+\sigma^{(b)}\tanh(\beta w)}{2}, 
 \label{eq:RS_m} \\
 \fl \hat{m}_*\left( \vec{\sigma}, \vec{s} \right) &=& \hat{\alpha} \int d\hat{\rho}\left(\hat{h} \right) d\hat{\mu} \left(\hat{w}|\hat{h}, s^{(1)} \right) e^{\sum_{a=1}^m \beta^\prime \hat{h} s^{(a)}} e^{\sum_{b=1}^n \beta \hat{w} \sigma^{(b)}},
 \label{eq:RS_mh}
\end{eqnarray}
where the functions $\rho \left(h \right)$, $\mu \left(w|h, s^{(1)} \right)$, $\hat{\rho}\left(\hat{h} \right)$, and $\hat{\mu} \left(\hat{w}|\hat{h}, s^{(1)} \right)$ are well-defined probability density functions.
Here, $\alpha$ and $\hat{\alpha}$ are constants and should be determined using a saddle-point condition.
Substituting (\ref{eq:RS_m}) and (\ref{eq:RS_mh}) into $-\beta g_{n,m}$ and optimizing with respect to $\alpha$ and $\hat{\alpha}$, we have
\begin{eqnarray}
 \fl -\beta g_{n,m} &=& \frac{C}{p} \log 2^{m+n} \cosh^m\left( \beta^\prime J \right) \cosh^n\left( \beta J \right) \sum_{s_{(1)}^{(1)}} \cdots \sum_{s_{(p)}^{(1)}} \int \prod_{j=1}^p d\rho\left(h_j \right) d\mu\left(w_j|h_j, s_{(j)}^{(1)} \right) \nonumber \\
 \fl && \qquad \times \mathbb{E}_J \left[ \frac{1+\tanh\left( \beta^\prime J \right)\prod_{j=1}^p s_{(j)}^{(1)}}{2} \prod_{j=1}^p \frac{1+\tanh\left( \beta^\prime h_j \right)s_{(j)}^{(1)}}{2} \right. \nonumber \\
 \fl && \qquad \times \left\{ \frac{1+\tanh\left( \beta^\prime J \right)\prod_{j=1}^p \tanh\left( \beta^\prime h_j \right)}{2} \right\}^{m-1} \nonumber \\
 \fl && \qquad \times \left. \left\{ \frac{1+\tanh\left( \beta J \right)\prod_{j=1}^p \tanh\left( \beta w_j \right)}{2} \right\}^{n} \right] \nonumber \\
 \fl && - C\log \sum_{s^{(1)}} \int d\rho \left(h \right) d\mu\left(w|h, s^{(1)} \right) \int d\hat{\rho}\left(\hat{h} \right) d\hat{\mu}\left(\hat{w}|\hat{h}, s^{(1)} \right) \nonumber \\
 \fl && \qquad \times \frac{1+\tanh\left( \beta^\prime h + \beta^\prime \hat{h} \right)s^{(1)}}{2} \nonumber \\
 \fl && \qquad \times \left\{ \frac{\cosh\left( \beta^\prime h + \beta^\prime \hat{h} \right)}{\cosh\left( \beta^\prime h \right)} \right\}^m \left\{ \frac{\cosh\left( \beta w + \beta \hat{w} \right)}{\cosh\left( \beta w \right)} \right\}^n \nonumber \\
 \fl && + \log \sum_{s^{(1)}} \int \prod_{d=1}^C d\hat{\rho}(\hat{h}_d) d\hat{\mu}\left(\hat{w}_d|\hat{h}_d, s^{(1)} \right) e^{\beta^\prime \left( h_0 + \sum_{d=1}^C \hat{h}_d \right)s^{(1)}} \nonumber \\
 \fl && \qquad \times \left\{ 2\cosh\left( \beta^\prime h_0 + \beta^\prime \sum_{d=1}^C \hat{h}_d \right) \right\}^{m-1} \nonumber \\
 \fl && \qquad \times \left\{ 2\cosh\left( \beta h_0 + \beta h_\mathrm{ext} s^{(1)} + \beta \sum_{d=1}^C \hat{w}_d \right) \right\}^n.
\end{eqnarray}
Now we are ready to perform the analytic continuation, and the Legendre transform of the Franz-Parisi potential is calculated:
\begin{eqnarray}
 \fl - \beta g(\beta, \beta^\prime, h_\mathrm{ext}) &=& \log 2 + \frac{C}{p} \mathbb{E}_J \left[ \log \cosh( \beta J ) \right] \nonumber \\
 \fl && + \sum_{s}\int \prod_{d=1}^C d\hat{\rho}(\hat{h}_d) d\hat{\mu}\left(\hat{w}_d|\hat{h}_d, s \right) \frac{1+s\tanh \left( \beta^\prime h_{0} + \beta^\prime \sum_{d=1}^C \hat{h}_d \right)}{2} \nonumber \\
 \fl && \qquad \times \log \cosh \left( \beta h_{0} + \beta h_\mathrm{ext} s + \beta \sum_{d=1}^C \hat{w}_d \right) \nonumber \\
 \fl && - C \sum_s \int d\rho \left(h \right) d\mu\left(w|h, s \right) \int d\hat{\rho}\left(\hat{h} \right) d\hat{\mu}\left(\hat{w}|\hat{h}, s \right) \nonumber \\
 \fl && \qquad \times \frac{1+s\tanh \left( \beta^\prime h + \beta^\prime \hat{h} \right)}{2} \nonumber \\
 \fl && \qquad \times \left[ \log \cosh(\beta w + \beta \hat{w}) - \log \cosh(\beta w) \right] \nonumber \\
 \fl && + \frac{C}{p} \sum_{s_{(1)}} \cdots \sum_{s_{(p)}} \int \prod_{j=1}^p d\rho\left(h_j \right) d\mu\left(w_j|h_j, s_{(j)} \right) \nonumber \\
 \fl && \qquad \times \mathbb{E}_J \left[ \frac{1+\tanh(\beta^\prime J)\prod_{j=1}^ps_{(j)}}{1+\tanh(\beta^\prime J)\prod_{j=1}^p\tanh\left( \beta^\prime h_j \right)} \right. \nonumber \\
 \fl && \qquad \times \prod_{j=1}^p \frac{1+s_{(j)}\tanh\left( \beta^\prime h_j \right)}{2} \nonumber \\
 \fl && \qquad \left. \times \log \left( 1 + \tanh(\beta J)\prod_{j=1}^p\tanh(\beta w_j)  \right) \right].
 \label{eq:FPL_final}
\end{eqnarray}

\subsection{Saddle-point equations}
Saddle-point equations are obtained by differentiating (\ref{eq:FPL_final}) with respect to $\mu \left(w|h, s \right)$ and $\hat{\mu} \left(\hat{w}|\hat{h}, s \right)$, and they are expressed as
\begin{eqnarray}
 \fl \hat{\rho}\left(\hat{h} \right) \hat{\mu} \left(\hat{w}|\hat{h}, s \right) &=& \sum_{s_{(1)}} \cdots \sum_{s_{(p-1)}} \int \prod_{j=1}^{p-1}d\rho\left(h_j \right) d\mu\left(w_j|h_j, s_{(j)} \right) \nonumber \\
 \fl && \qquad \times \mathbb{E}_J \left[ \frac{1+\tanh(\beta^\prime J)s\prod_{j=1}^{p-1}s_{(j)}}{1+\tanh(\beta^\prime J)s\prod_{j=1}^{p-1}\tanh\left( \beta^\prime h_j \right)} \right. \nonumber \\
 \fl && \qquad \times \prod_{j=1}^{p-1} \frac{1+s_{(j)}\tanh\left( \beta^\prime h_j \right)}{2} \nonumber \\
 \fl && \qquad \times \delta \left( \hat{h} - \frac{1}{\beta^\prime}\tanh^{-1}\left( \tanh(\beta^\prime J)\prod_{j=1}^{p-1} \tanh(\beta^\prime h_j) \right)  \right) \nonumber \\
 \fl && \qquad \left. \times \delta \left( \hat{w} - \frac{1}{\beta}\tanh^{-1}\left( \tanh(\beta J)\prod_{j=1}^{p-1} \tanh(\beta w_j) \right)  \right) \right],
 \label{eq:DE_Ph_wf} \\
 \fl \rho \left(h \right) \mu \left(w|h, s \right) &=& \int \prod_{d=1}^{C-1} d\hat{\rho}\left(\hat{h}_d \right) d\hat{\mu}\left(\hat{w}_d|\hat{h}_d, s \right) \delta\left( h - h_{0} - \sum_{d=1}^{C-1} \hat{h}_d \right) \nonumber \\
 \fl && \qquad \times \delta\left( w - h_{0} - h_\mathrm{ext} s - \sum_{d=1}^{C-1} \hat{w}_d \right).
 \label{eq:DE_P_wf}
\end{eqnarray}
When we integrate these self-consistent equations with respect to $\hat{w}$ and $w$, respectively, we produce standard RS self-consistent equations.
Therefore, we obtain the RS expression of the Legendre transform of the Franz-Parisi potential for the $p$-spin model on regular random graphs under a uniform magnetic field.
These results indeed coincide with those obtained by using the cavity method \cite{ZdeKrz2010}.
Therefore, our results provide a basis for the results obtained using the cavity method in terms of the replica method.

Our results can also be extended to other graphs, such as Erd\"{o}s-R\'{e}nyi graphs, in a straightforward manner.
It should be noted that when $h_0=0$, we have $\rho \left( h \right)=\delta\left(h\right)$ and $\hat{\rho}\left( \hat{h} \right)=\delta\left( \hat{h} \right)$, and the probability distributions are respectively simplified to $\mu \left(w|0, s \right)=P\left( sw \right)$ and $\hat{\mu} \left(\hat{w}|0, s \right)=\hat{P}\left( s\hat{w} \right)$ because of the spin reversal symmetry.

\section{Ultrametric structures}
\label{sec:1rsb}
\begin{figure}[tbp]
\begin{center}
\includegraphics{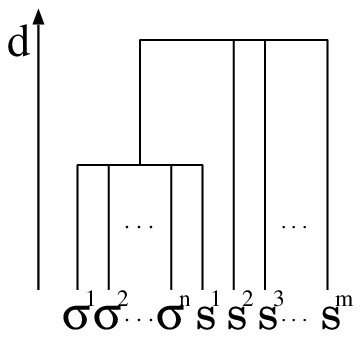}
\caption{Ultrametric structure that appears at extrema of Franz-Parisi potential. 
For $s^{(a)}$ $(a=1,2,\cdots,m)$ and $\sigma^{(b)}$ $(b=1,2,\cdots,n)$, let us define a ``distance'' to be  $d(x,y)\equiv 1-\mathbb{E}\left[xys^{(a_1)}s^{(a_2)}\cdots s^{(a_{k-1})}\sigma^{(b_1)}\sigma^{(b_2)}\cdots \sigma^{(b_l)}\right]$, where $x$ and $y$ are elements of $s^{(a)}$ and $\sigma^{(b)}$ that differ from any of $s^{(a_1)}$, $s^{(a_2)}$, $\cdots$, $s^{(a_{k-1})}$ and $\sigma^{(b_1)}$, $\sigma^{(b_2)}$, $\cdots$, $\sigma^{(b_l)}$. 
This structure indicates that the ultrametric relation $d(s^{(1)},\sigma^{(b)}) \leq d(s^{(1)},s^{(a)})=d(\sigma^{(b)},s^{(a)})$ $(a\neq 1)$ holds, which gives us (\ref{eq:ultra}).}
\label{fig:1RSBstructure}
\end{center}
\end{figure}

We now prove that the above formalism is equivalent to the 1RSB cavity method with $x=1$.
Let us consider the case $\beta^\prime=\beta$ and $h_\mathrm{ext}=0$.
Analogous to the results for fully connected models, we expect $s^{(1)}$ to behave similarly to $\sigma^{(b)}$, as well as to $s^{(a)}$ \cite{FP1995, UedSas2014}.
In other words, we expect the averaged multi-body overlaps $\mathbb{E}\left[\left\langle N^{-1}\sum_{i=1}^Ns_i^{(1)}s_i^{(a_1)}\cdots s_i^{(a_k)} \sigma_i^{(b_1)}\cdots \sigma_i^{(b_l)}\right\rangle\right]$ and $\mathbb{E}\left[\left\langle N^{-1}\sum_{i=1}^Ns_i^{(a_1)}\cdots s_i^{(a_k)} \sigma_i^{(b_1)}\cdots \sigma_i^{(b_{l+1})}\right\rangle\right]$ to be equivalent for arbitrary $k$ and $l$.
This expectation implies that the ultrametric structure represented in Figure \ref{fig:1RSBstructure} appears at the extrema of the Franz-Parisi potential.
Averages of multi-body overlaps with respect to cavity field distributions can be written as
\begin{eqnarray}
 r_{k,l} &\equiv& \sum_s \int d\rho\left(h \right) d\mu\left(w|h, s \right) \frac{e^{\beta h s}}{2\cosh(\beta h)} s \tanh^k(\beta h) \tanh^l(\beta w), \\
 q_{k,l+1} &\equiv& \sum_s \int d\rho\left(h \right) d\mu\left(w|h, s \right) \frac{e^{\beta h s}}{2\cosh(\beta h)} \tanh^k(\beta h) \tanh^{l+1}(\beta w).
\end{eqnarray}
With this, we can prove the existence of a solution that satisfies the relation
\begin{eqnarray}
 r_{k,l}=q_{k,l+1}\qquad (\forall k,l\geq 0),
 \label{eq:ultra}
\end{eqnarray}
which implies the ultrametric structure. 

In order to simplify the notation, we define cavity fields $\tilde{h}$ and $\tilde{w}$ to be $h=h_0+\tilde{h}$ and $w=h_0+\tilde{w}+\tilde{h}$, respectively.
Defining the distributions of $\tilde{h}$ and $\tilde{w}$ by $\tilde{\rho}\left( \tilde{h} \right)$ and $\tilde{\mu}\left( \tilde{w}|\tilde{h}, s \right)$, respectively, self-consistent equations for these distributions can be written:
\begin{eqnarray}
 \fl \hat{\rho}\left(\hat{h} \right) \hat{\mu} \left(\hat{w}|\hat{h}, s \right) &=& \sum_{s_{(1)}} \cdots \sum_{s_{(p-1)}} \int \prod_{j=1}^{p-1}d\tilde{\rho}\left(\tilde{h}_j \right) d\tilde{\mu}\left(\tilde{w}_j|\tilde{h}_j, s_{(j)} \right) \nonumber \\
 \fl && \qquad \times \mathbb{E}_J \left[ \frac{1+\tanh(\beta J)s\prod_{j=1}^{p-1}s_{(j)}}{1+\tanh(\beta J)s\prod_{j=1}^{p-1}\tanh\left( \beta \tilde{h}_j + \beta h_0 \right)} \right. \nonumber \\
 \fl && \qquad \times \prod_{j=1}^{p-1} \frac{1+s_{(j)}\tanh\left( \beta \tilde{h}_j + \beta h_0 \right)}{2} \nonumber \\
 \fl && \qquad \times \delta \left( \hat{h} - \frac{1}{\beta}\tanh^{-1}\left( \tanh(\beta J)\prod_{j=1}^{p-1} \tanh(\beta \tilde{h}_j + \beta h_0) \right)  \right) \nonumber \\
 \fl && \qquad \times \delta \left( \hat{w} - \frac{1}{\beta}\tanh^{-1}\left( \tanh(\beta J) \right. \right. \nonumber \\
 \fl && \qquad \times \left. \left. \left. \prod_{j=1}^{p-1} \tanh(\beta \tilde{w}_j + \beta \tilde{h}_j + \beta h_0) \right)  \right) \right],
 \label{eq:DEmod_Ph_wf} \\
 \fl \tilde{\rho} \left(\tilde{h} \right) \tilde{\mu}\left( \tilde{w}|\tilde{h}, s \right) &=& \int \prod_{d=1}^{C-1} d\hat{\rho}\left(\hat{h}_d \right) d\hat{\mu}\left(\hat{w}_d|\hat{h}_d, s \right) \delta\left( \tilde{h} - \sum_{d=1}^{C-1} \hat{h}_d \right) \nonumber \\
 \fl && \qquad \times \delta\left( \tilde{w} + \tilde{h} - \sum_{d=1}^{C-1} \hat{w}_d \right).
 \label{eq:DEmod_P_wf}
\end{eqnarray}
Furthermore, the multi-body overlaps can be expressed as
\begin{eqnarray}
 \fl r_{k,l} &=& \sum_s \int d\tilde{\rho}\left(\tilde{h} \right) d\tilde{\mu}\left( \tilde{w}|\tilde{h}, s \right) \frac{e^{\beta \left(\tilde{h}+h_0\right) s}}{2\cosh\left(\beta \tilde{h} + \beta h_0\right)} s \tanh^k\left(\beta \tilde{h} + \beta h_0\right) \nonumber \\
 \fl && \qquad \times \tanh^l\left(\beta \tilde{w} + \beta \tilde{h} + \beta h_0\right),
 \label{eq:rtilde} \\
 \fl q_{k,l+1} &=& \sum_s \int d\tilde{\rho}\left(\tilde{h} \right) d\tilde{\mu}\left( \tilde{w}|\tilde{h}, s \right) \frac{e^{\beta \left(\tilde{h}+h_0\right) s}}{2\cosh\left(\beta \tilde{h} + \beta h_0\right)} \tanh^k\left(\beta \tilde{h} + \beta h_0\right) \nonumber \\
 \fl && \qquad \times \tanh^{l+1}\left(\beta \tilde{w} + \beta \tilde{h} + \beta h_0\right).
 \label{eq:qtilde}
\end{eqnarray}
Hence, the relation (\ref{eq:ultra}) implies the equivalency of the right-hand sides of (\ref{eq:rtilde}) and (\ref{eq:qtilde}).
We will prove that they indeed become equivalent under the relation (\ref{eq:ultra}).

As demonstrated in \ref{ap:ultra}, we can prove
\begin{eqnarray}
 \tilde{\rho} \left(\tilde{h} \right) \tilde{\mu}\left( \tilde{w}|\tilde{h}, 1 \right) e^{-2\beta \tilde{w}} = \tilde{\rho} \left(\tilde{h} \right) \tilde{\mu}\left( \tilde{w}|\tilde{h}, -1 \right)
 \label{eq:symm}
\end{eqnarray}
under the relation (\ref{eq:ultra}).
This means that the dependence of $\tilde{\mu}\left( \tilde{w}|\tilde{h}, s \right)$ on $s$ is limited: 
\begin{eqnarray}
 \tilde{\mu}\left( \tilde{w}|\tilde{h}, s \right) = \tilde{\phi}\left( \tilde{h}, \tilde{w} \right) e^{\beta \tilde{w} s}
 \label{eq:ultra_cavityt}
\end{eqnarray}
with some function $\tilde{\phi}$.
Applying this relation to (\ref{eq:rtilde}) and (\ref{eq:qtilde}) gives us
\begin{eqnarray}
 \fl r_{k,l} &=& \int d\tilde{h} d\tilde{w} \tilde{\rho} \left(\tilde{h} \right) \tilde{\phi}\left( \tilde{h}, \tilde{w} \right) \frac{\sinh\left(\beta \tilde{w} + \beta \tilde{h} + \beta h_0\right)}{\cosh\left(\beta \tilde{h} + \beta h_0\right)} \tanh^k\left(\beta \tilde{h} + \beta h_0\right) \nonumber \\
 \fl && \qquad \times \tanh^l\left(\beta \tilde{w} + \beta \tilde{h} + \beta h_0\right), \\
 \fl q_{k,l+1} &=& \int d\tilde{h} d\tilde{w} \tilde{\rho} \left(\tilde{h} \right) \tilde{\phi}\left( \tilde{h}, \tilde{w} \right) \frac{\cosh\left(\beta \tilde{w} + \beta \tilde{h} + \beta h_0\right)}{\cosh\left(\beta \tilde{h} + \beta h_0\right)} \tanh^k\left(\beta \tilde{h} + \beta h_0\right) \nonumber \\
 \fl && \qquad \times \tanh^{l+1}\left(\beta \tilde{w} + \beta \tilde{h} + \beta h_0\right).
\end{eqnarray}
These results, in conjunction with the identity $\sinh(\cdots)=\tanh(\cdots)\cosh(\cdots)$, indicate that $r_{k,l}$ and $q_{k,l+1}$ are equivalent.
Thus, the existence of the solution that satisfies (\ref{eq:ultra}) has been proved.
We note that this result is a generalization of the relations $r_{0,1}=q_{0,2}$ and $r_{1,0}=q_{1,1}$ for the fully connected model.

Similarly, under the ultrametric structure (\ref{eq:ultra}), we can prove the relation
\begin{eqnarray}
 \hat{\mu}\left( \hat{w}|\hat{h}, s \right) = \hat{\phi}\left( \hat{h}, \hat{w} \right) e^{\beta (\hat{w}-\hat{h}) s}.
 \label{eq:ultra_cavityh}
\end{eqnarray}
Therefore, if we define the averages of the multi-body overlaps with respect to the local field distributions to be
\begin{eqnarray}
 \fl R_{k,l} &\equiv& \sum_s \int \prod_{d=1}^{C} d\hat{\rho}\left(\hat{h}_d \right) d\hat{\mu}\left(\hat{w}_d|\hat{h}_d, s \right) \frac{e^{\beta \left(\sum_{d=1}^{C} \hat{h}_d+h_0\right) s}}{2\cosh\left(\beta \sum_{d=1}^{C} \hat{h}_d + \beta h_0\right)} s \nonumber \\
 \fl && \qquad \times \tanh^k\left(\beta \sum_{d=1}^{C} \hat{h}_d + \beta h_0\right) \tanh^l\left(\beta \sum_{d=1}^{C} \hat{w}_d + \beta h_0\right), \\
 \fl Q_{k,l+1} &\equiv& \sum_s \int \prod_{d=1}^{C} d\hat{\rho}\left(\hat{h}_d \right) d\hat{\mu}\left(\hat{w}_d|\hat{h}_d, s \right) \frac{e^{\beta \left(\sum_{d=1}^{C} \hat{h}_d+h_0\right) s}}{2\cosh\left(\beta \sum_{d=1}^{C} \hat{h}_d + \beta h_0\right)} \nonumber \\
 \fl && \qquad \times \tanh^k\left(\beta \sum_{d=1}^{C} \hat{h}_d + \beta h_0\right) \tanh^{l+1}\left(\beta \sum_{d=1}^{C} \hat{w}_d + \beta h_0\right),
\end{eqnarray}
our solution also satisfies the relation
\begin{eqnarray}
 R_{k,l}=Q_{k,l+1} \qquad  (\forall k,l\geq 0).
\end{eqnarray}
It should be noted that the relation in (\ref{eq:ultra_cavityt}) is rewritten as 
\begin{eqnarray}
 \mu\left( w|h, s \right) = \phi\left( h, w \right) e^{\beta (w-h) s}
 \label{eq:ultra_cavity}
\end{eqnarray}
in the original notation.

Finally, we show that our solution at the extrema and the 1RSB solution with $x=1$ are equivalent.
In order to do this, we need to define the distributions
\begin{eqnarray}
 P_2\left( h, w \right) &\equiv& \rho \left(h \right) \phi\left( h, w \right) \frac{\cosh\left( \beta w \right)}{\cosh\left( \beta h \right)}, \\
 \hat{P}_2\left( \hat{h}, \hat{w} \right) &\equiv& \hat{\rho} \left(\hat{h} \right) \hat{\phi}\left( \hat{h}, \hat{w} \right) \frac{\cosh\left( \beta \hat{w} \right)}{\cosh\left( \beta \hat{h} \right)}.
\end{eqnarray}
Substituting (\ref{eq:ultra_cavityh}) and (\ref{eq:ultra_cavity}) into (\ref{eq:DE_Ph_wf}) and (\ref{eq:DE_P_wf}) and using these definitions let us determine that $P_2$ and $\hat{P}_2$ satisfy the respective self-consistent equations
\begin{eqnarray}
 \fl \hat{P}_2\left( \hat{h}, \hat{w} \right) &=& \int \prod_{j=1}^{p-1} dP_2\left( h_j, w_j \right) \mathbb{E}_J \left[ \delta \left( \hat{h} - \frac{1}{\beta}\tanh^{-1}\left( \tanh(\beta J)\prod_{j=1}^{p-1} \tanh(\beta h_j) \right)  \right) \right. \nonumber \\
 \fl && \qquad \left. \times \delta \left( \hat{w} - \frac{1}{\beta}\tanh^{-1}\left( \tanh(\beta J)\prod_{j=1}^{p-1} \tanh(\beta w_j) \right)  \right) \right], \\
 \fl P_2\left( h, w \right) &=& \int \prod_{d=1}^{C-1} d\hat{P}_2\left( \hat{h}_d, \hat{w}_d \right) \frac{\cosh\left( \beta h_0 + \beta \sum_{d=1}^{C-1} \hat{w}_{d} \right)}{\prod_{d=1}^{C-1} \cosh\left( \beta \hat{w}_{d} \right)} \frac{\prod_{d=1}^{C-1} \cosh\left( \beta \hat{h}_{d} \right)}{\cosh\left( \beta h_0 + \beta \sum_{d=1}^{C-1} \hat{h}_{d} \right)} \nonumber \\
 \fl && \qquad \times \delta\left( h - h_{0} - \sum_{d=1}^{C-1} \hat{h}_d \right) \delta\left( w - h_{0} - \sum_{d=1}^{C-1} \hat{w}_d \right).
\end{eqnarray}
These equations are equivalent to the 1RSB cavity equations with Parisi parameter $x=1$ \cite{MRS2008, Zde2009}.
Furthermore, at the extrema, the Legendre transform of the Franz-Parisi potential can be expressed using these distributions by
\begin{eqnarray}
 \fl - \beta g(\beta, \beta, 0) &=& \log 2 + \frac{C}{p} \mathbb{E}_J \left[ \log \cosh( \beta J ) \right] \nonumber \\
 \fl && + \int \prod_{d=1}^C d\hat{P}_2\left( \hat{h}_d, \hat{w}_d \right) \frac{\cosh\left( \beta h_0 + \beta \sum_{d=1}^{C} \hat{w}_{d} \right)}{\prod_{d=1}^{C} \cosh\left( \beta \hat{w}_{d} \right)} \nonumber \\
 \fl && \qquad \times \frac{\prod_{d=1}^{C} \cosh\left( \beta \hat{h}_{d} \right)}{\cosh\left( \beta h_0 + \beta \sum_{d=1}^{C} \hat{h}_{d} \right)} \log \cosh\left( \beta h_0 + \beta \sum_{d=1}^{C} \hat{w}_{d} \right) \nonumber \\
 \fl && - C\int dP_2\left( h, w \right) d\hat{P}_2\left( \hat{h}, \hat{w} \right) \frac{1+\tanh\left( \beta w \right)\tanh\left( \beta \hat{w} \right)}{1+\tanh\left( \beta h \right)\tanh\left( \beta \hat{h} \right)} \nonumber \\
 \fl && \qquad \times \left[ \log \cosh\left( \beta w + \beta \hat{w} \right) - \log \cosh\left( \beta w \right) \right] \nonumber \\
 \fl && + \frac{C}{p} \int \prod_{j=1}^{p} dP_2\left( h_j, w_j \right) \mathbb{E}_J \left[ \frac{1+\tanh(\beta J)\prod_{j=1}^{p} \tanh(\beta w_j)}{1+\tanh(\beta J)\prod_{j=1}^{p} \tanh(\beta h_j)} \right. \nonumber \\
 \fl && \qquad \left. \times \log \left( 1+\tanh(\beta J)\prod_{j=1}^{p} \tanh(\beta w_j) \right) \right].
 \label{eq:ntlm}
\end{eqnarray}
In particular, the trivial solution $P_2\left( h, w \right)=\rho(h)\delta(w-h)$ and $\hat{P}_2\left( \hat{h}, \hat{w} \right)=\hat{\rho}\left( \hat{h} \right)\delta\left( \hat{w}-\hat{h} \right)$ gives us
\begin{eqnarray}
 \fl - \beta g_\mathrm{para}(\beta, \beta, 0) &=& \log 2 + \frac{C}{p} \mathbb{E}_J \left[ \log \cosh( \beta J ) \right] \nonumber \\
 \fl && + \int \prod_{d=1}^C d\hat{\rho}\left( \hat{w}_d \right) \log \cosh\left( \beta h_0 + \beta \sum_{d=1}^{C} \hat{w}_{d} \right) \nonumber \\
 \fl && - C\int d\rho\left( w \right) d\hat{\rho}\left( \hat{w} \right) \left[ \log \cosh\left( \beta w + \beta \hat{w} \right) - \log \cosh\left( \beta w \right) \right] \nonumber \\
 \fl && + \frac{C}{p} \int \prod_{j=1}^{p} d\rho\left( w_j \right) \mathbb{E}_J \left[ \log \left( 1+\tanh(\beta J)\prod_{j=1}^{p} \tanh(\beta w_j) \right) \right],
 \label{eq:tlm}
\end{eqnarray}
which is equivalent to the replica symmetric free energy.

In the 1RSB cavity method, a free energy of the system is obtained by optimizing the 1RSB free energy $f_\mathrm{1RSB}(x)$ with respect to Parisi parameter $x$.
At the 1RSB transition temperature, the optimal value of $x$ is $x=1$.
Therefore, the transition temperature is calculated using the condition $\left.\partial f_\mathrm{1RSB}/\partial x \right|_{x=1}=0$.
Let us recall that physical quantities at $x=1$ can be expressed only by $P_2$ and $\hat{P}_2$ without using functional distributions \cite{MRS2008, Zde2009}.
We see that the 1RSB transition condition given in previous studies is equivalent to the phase transition condition in (\ref{eq:FP_transition}) with (\ref{eq:ntlm}) and (\ref{eq:tlm}).
Thus, we have proved that the calculation of the Legendre transform of the Franz-Parisi potential under the RS ansatz with the ultrametric structure (\ref{eq:ultra}) is equivalent to the 1RSB cavity method with $x=1$.
Our results can be straightforwardly generalized to other models.

It should be noted that this equivalence was also discussed in \cite{ZdeKrz2010}.
There, the authors started from the 1RSB cavity equations with $x=1$ for $P_2\left( h, w \right)$ and $\hat{P}_2\left( \hat{h}, \hat{w} \right)$ and then defined spin-dependent probability distributions as (\ref{eq:ultra_cavity}) and (\ref{eq:ultra_cavityh}), which are shown to satisfy the self-consistent equations (\ref{eq:DE_Ph_wf}) and (\ref{eq:DE_P_wf}).
In contrast, our result could be considered to be a reinterpretation of the equivalence in terms of the ultrametric structure (\ref{eq:ultra}) that appears at the extrema of the Franz-Parisi potential of $\beta^\prime=\beta$. 

As a special case, let us consider $h_0=0$.
In this case, (\ref{eq:ultra_cavity}) and (\ref{eq:ultra_cavityh}) are $P\left( w \right)e^{-2\beta w}= P\left( -w \right)$ and $\hat{P}\left( \hat{w} \right)e^{-2\beta \hat{w}}= \hat{P}\left( -\hat{w} \right)$, respectively.
This case was also analyzed in \cite{ZdeKrz2010} using gauge theory as a basis.

\section{Numerical analysis}
\label{sec:numerical}
Table \ref{tab:moments} displays the numerical results for the first nine moments $r_{k,l}$ and $q_{k,l+1}$ that were calculated using the self-consistent equations (\ref{eq:DE_Ph_wf}) and (\ref{eq:DE_P_wf}).
\begin{table}[tb]
  \begin{center}
   \begin{tabular}{c c c}
    \hline
    $(k,l)$ & $r_{k,l}$ & $q_{k,l+1}$\\
    \hline
    (0,0) & 0.351 & 0.351\\
    (0,1) & 0.854 & 0.854\\
    (0,2) & 0.343 & 0.343\\
    (1,0) & 0.171 & 0.171\\
    (1,1) & 0.303 & 0.303\\
    (1,2) & 0.166 & 0.166\\
    (2,0) & 0.088 & 0.088\\
    (2,1) & 0.148 & 0.148\\
    (2,2) & 0.086 & 0.086\\
    \hline
   \end{tabular}
   \caption{Moments $r_{k,l}$ and $q_{k,l+1}$ for $p=3$, $C=4$, $h_0=0.30$ and $T=0.77$}
   \label{tab:moments}
 \end{center}
\end{table}
These results indicate that the ultrametric structure (\ref{eq:ultra}) holds.
We should note that the self-consistent equations (\ref{eq:DE_Ph_wf}) and (\ref{eq:DE_P_wf}) can converge to solutions that do not satisfy the relation in (\ref{eq:ultra}), depending on the initial condition.
Therefore, (\ref{eq:ultra}) is a useful means to check the convergence of population dynamics to the 1RSB-type solution.

\begin{figure}[tbp]
\begin{center}
\includegraphics{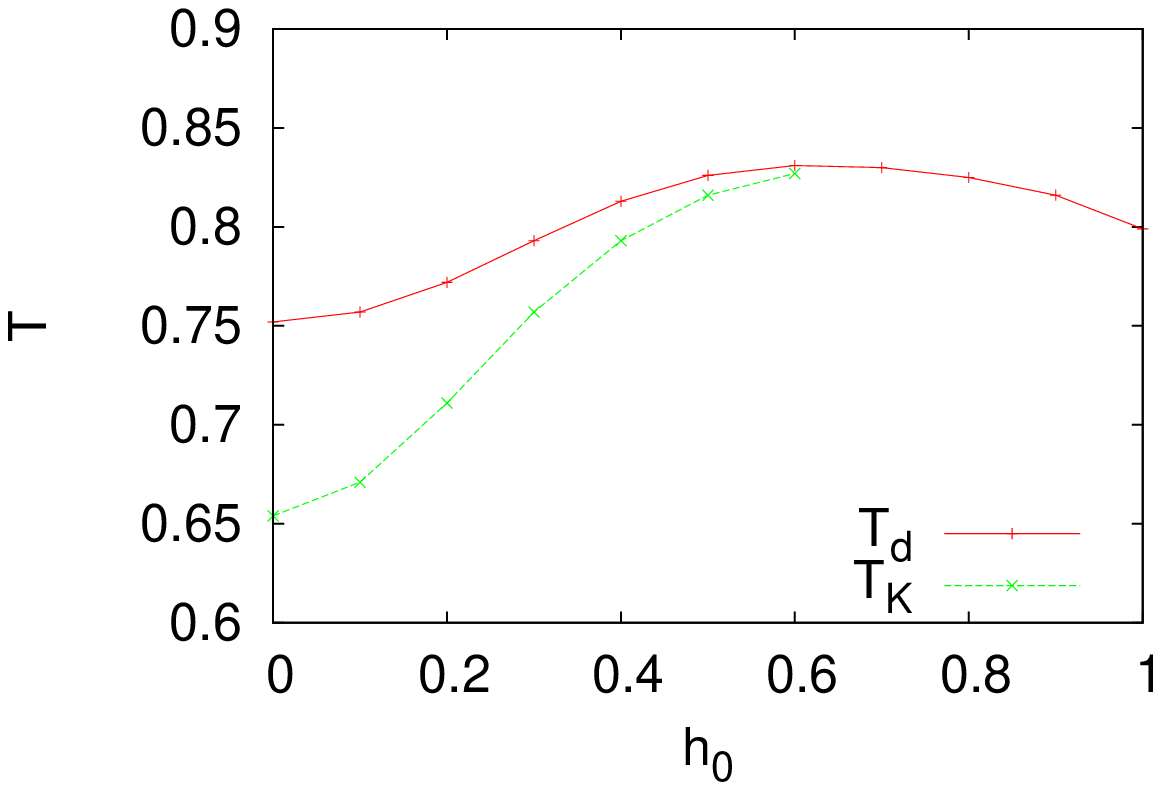}
\caption{Transition temperature when $p=3$ and $C=4$}
\label{fig:FPwithfield_transition}
\end{center}
\end{figure}

In Figure \ref{fig:FPwithfield_transition}, we also provide the results for the static 1RSB temperature $T_\mathrm{K}$ that was calculated using the phase transition condition in (\ref{eq:FP_transition}) and the dynamical 1RSB temperature $T_\mathrm{d}$ that was calculated using the appearance of a nontrivial solution for the self-consistent equations (\ref{eq:DE_Ph_wf}) and (\ref{eq:DE_P_wf}).
The shape of this graph is similar to the one obtained for the fully connected model by using the Franz-Parisi potential with the RS ansatz.
In particular, there is a critical point $(h_{0\mathrm{c}}, T_\mathrm{c})$ in which $T_\mathrm{d}=T_\mathrm{K}$.
In the region $h_0\geq h_{0\mathrm{c}}$, overlap continuously increases at the transition temperature.
A similar phase diagram was also created for the fully connected spherical $p$-spin model \cite{CriSom1992}.
It should be noted that because the thermodynamic value of $x$ is generally not $x=1$ in the region $h_0>h_{0\mathrm{c}}$, we need to perform the standard 1RSB analysis in order to accurately examine this region.

\section{Conclusion}
\label{sec:conclusion}
In this paper, we proposed a method for calculating the Franz-Parisi potential for spin glass models on sparse random graphs using the replica method under the replica symmetric ansatz.
In order to do this, we first introduced cavity fields, which require the spin variable $s^{(1)}$ to connect the $\vec{\sigma}$ and $\vec{s}$ systems.
Our results are reinterpretations of those obtained using the cavity method.
We also proved that the self-consistent equations have a solution with the characteristic structure in (\ref{eq:ultra}) for multi-body overlaps when $\beta^\prime=\beta$ and $h_\mathrm{ext}=0$ and that, under this structure, the self-consistent equations are equivalent to the 1RSB cavity equation with Parisi parameter $x=1$.
The relation in (\ref{eq:ultra}) is useful for checking whether the population dynamics for equations (\ref{eq:DE_Ph_wf}) and (\ref{eq:DE_P_wf}) correctly converge to the 1RSB-type solution.

Furthermore, we exhibited the results obtained by applying our method to the calculation of the transition temperature of the $p$-spin model on regular random graphs under a uniform magnetic field.
Similar to the $x=1$ 1RSB cavity method, the transition temperature was calculated using a computational cost as low as that used in the RS cavity method.
The obtained phase diagram is similar to that of the fully connected $p$-spin model under a uniform magnetic field.
Because we cannot correctly calculate transition temperatures using our method in the region $x\neq 1$, it should be noted that we must perform the standard 1RSB analysis for this region.
In future work, we should also compare our numerical results with those obtained using other methods, such as the Monte Carlo method.

\ack
We would like to thank S. Takabe for the valuable discussions.
This work was supported by a Grant-in-Aid for JSPS Fellows (No. 2681) (MU), KAKENHI No. 25120013 (YK), and the JSPS Core-to-Core Program ``Nonequilibrium dynamics of soft matter and information.''

\appendix

\section{Franz-Parisi potential}
\label{ap:potential}
In this appendix, we propose a method for analyzing the Franz-Parisi potential under the RS ansatz.
The Franz-Parisi potential is defined by (\ref{eq:FP_potential}).
Using a calculation similar to that used in $-\beta g(\beta, \beta^\prime, h_\mathrm{ext})$, we derive for the $p$-spin model on regular random graphs under a uniform magnetic field:
\begin{eqnarray}
 \fl - \beta v(\beta, \beta^\prime, q) &=& \log 2 + \frac{C}{p} \mathbb{E} \left[ \log \cosh( \beta J ) \right] - \beta h_\mathrm{ext} q \nonumber \\
 \fl && + \sum_{s}\int \prod_{d=1}^C d\hat{\rho}(\hat{h}_d) d\hat{\mu}\left(\hat{w}_d|\hat{h}_d, s \right) \frac{1+s\tanh \left( \beta^\prime h_{0} + \beta^\prime \sum_{d=1}^C \hat{h}_d \right)}{2} \nonumber \\
 \fl && \qquad \times \log \cosh \left( \beta h_{0} + \beta h_\mathrm{ext} s + \beta \sum_{d=1}^C \hat{w}_d \right) \nonumber \\
 \fl && - C \sum_s \int d\rho \left(h \right) d\mu\left(w|h, s \right) \int d\hat{\rho}\left(\hat{h} \right) d\hat{\mu}\left(\hat{w}|\hat{h}, s \right) \nonumber \\
 \fl && \qquad \times \frac{1+s\tanh \left( \beta^\prime h + \beta^\prime \hat{h} \right)}{2} \nonumber \\
 \fl && \qquad \times \left[ \log \cosh(\beta w + \beta \hat{w}) - \log \cosh(\beta w) \right] \nonumber \\
 \fl && + \frac{C}{p} \sum_{s_{(1)}} \cdots \sum_{s_{(p)}} \int \prod_{j=1}^p d\rho\left(h_j \right) d\mu\left(w_j|h_j, s_{(j)} \right) \nonumber \\
 \fl && \qquad \times \mathbb{E} \left[ \frac{1+\tanh(\beta^\prime J)\prod_{j=1}^ps_{(j)}}{1+\tanh(\beta^\prime J)\prod_{j=1}^p\tanh\left( \beta^\prime h_j \right)} \right. \nonumber \\
 \fl && \qquad \times \prod_{j=1}^p \frac{1+s_{(j)}\tanh\left( \beta^\prime h_j \right)}{2} \nonumber \\
 \fl && \qquad \left. \times \log \left( 1 + \tanh(\beta J)\prod_{j=1}^p\tanh(\beta w_j)  \right) \right].
\end{eqnarray}
This gives us the saddle-point equations for the distributions:
\begin{eqnarray}
 \fl \hat{\rho}\left(\hat{h} \right) \hat{\mu} \left(\hat{w}|\hat{h}, s \right) &=& \sum_{s_{(1)}} \cdots \sum_{s_{(p-1)}} \int \prod_{j=1}^{p-1}d\rho\left(h_j \right) d\mu\left(w_j|h_j, s_{(j)} \right) \nonumber \\
 \fl && \qquad \times \mathbb{E}\left[ \frac{1+\tanh(\beta^\prime J)s\prod_{j=1}^{p-1}s_{(j)}}{1+\tanh(\beta^\prime J)s\prod_{j=1}^{p-1}\tanh\left( \beta^\prime h_j \right)} \right. \nonumber \\
 \fl && \qquad \times \prod_{j=1}^{p-1} \frac{1+s_{(j)}\tanh\left( \beta^\prime h_j \right)}{2} \nonumber \\
 \fl && \qquad \times \delta \left( \hat{h} - \frac{1}{\beta^\prime}\tanh^{-1}\left( \tanh(\beta J)\prod_{j=1}^{p-1} \tanh(\beta h_j) \right)  \right) \nonumber \\
 \fl && \qquad \left. \times \delta \left( \hat{w} - \frac{1}{\beta}\tanh^{-1}\left( \tanh(\beta J)\prod_{j=1}^{p-1} \tanh(\beta w_j) \right)  \right) \right], \\
 \fl \rho \left(h \right) \mu \left(w|h, s \right) &=& \int \prod_{d=1}^{C-1} d\hat{\rho}\left(\hat{h}_d \right) d\hat{\mu}\left(\hat{w}_d|\hat{h}_d, s \right) \delta\left( h - h_{0} - \sum_{d=1}^{C-1} \hat{h}_d \right) \nonumber \\
 \fl && \qquad \times \delta\left( w - h_{0} - h_\mathrm{ext} s - \sum_{d=1}^{C-1} \hat{w}_d \right).
\end{eqnarray}
Saddle-point condition for $h_\mathrm{ext}$ gives the equation
\begin{eqnarray}
 \fl q &=& \sum_s \int d\rho \left(h \right) d\mu \left(w|h, s \right) d\hat{\rho}\left(\hat{h} \right) d\hat{\mu} \left(\hat{w}|\hat{h}, s \right) \frac{1+s\tanh \left( \beta^\prime h + \beta^\prime \hat{h} \right)}{2} \nonumber \\
 \fl && \qquad \times s \tanh\left( \beta w + \beta \hat{w} \right).
\end{eqnarray}
This relation indicates that $h_\mathrm{ext}$ is self-consistently determined as a function of $q$.

\begin{figure}[tbp]
\begin{center}
\includegraphics{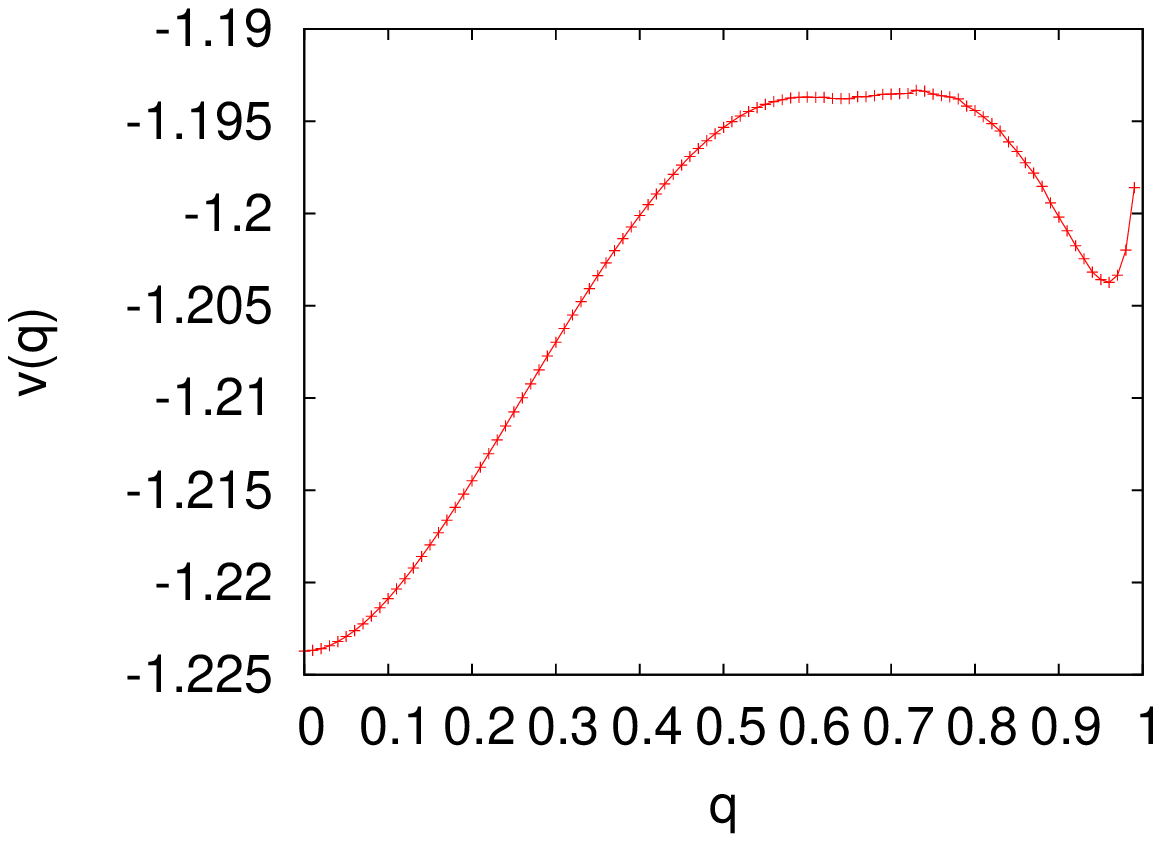}
\caption{Franz-Parisi potential for $p=3$, $C=4$, $h_0=0$, and $T=T^\prime=0.70$}
\label{fig:FPP_0.700}
\end{center}
\end{figure}

A graph of the Franz-Parisi potential for $h_0=0$ is displayed in Figure \ref{fig:FPP_0.700}.
Computation is performed using population dynamics with a fixed $q$, such as in \cite{SulZde2010}, and the partial update of the populations.
Even for sparse systems, we can check the existence of a potential barrier between a high-$q$ and low-$q$ phase similarly to the fully connected models.
It should be noted that because the RS solution may be unstable in the intermediate $q$ region, further analysis based on the 1RSB ansatz is required.

\section{Derivation of (\ref{eq:znmcal})}
\label{ap:znmcal}
In this appendix, we derive (\ref{eq:znmcal}).
By definition, $\mathbb{E} \left[ Z_{n,m} \right]$ can be rewritten as
\begin{eqnarray}
 \fl \mathbb{E} \left[ Z_{n,m} \right] &=& \sum_{\left\{ \textrm{\boldmath $s$}^{(a)} \right\}} \sum_{\left\{ \textrm{\boldmath $\sigma$}^{(b)} \right\}} e^{ \beta^\prime h_0 \sum_{a=1}^m \sum_{i=1}^N s_i^{(a)} + \beta h_0 \sum_{b=1}^n \sum_{i=1}^N \sigma_i^{(b)} + \beta h_\mathrm{ext} \sum_{b=1}^n \sum_{i=1}^N \sigma_i^{(b)} s_i^{(1)}} \nonumber \\
 \fl && \qquad \times \mathbb{E}_g \mathbb{E}_J \left[ e^{ \sum_{i_1< \cdots< i_p} g_{i_1, \cdots, i_p} J_{i_1, \cdots, i_p} \left( \beta^\prime \sum_{a=1}^m s_{i_1}^{(a)}\cdots s_{i_p}^{(a)} + \beta \sum_{b=1}^n \sigma_{i_1}^{(b)}\cdots \sigma_{i_p}^{(b)} \right) } \right] \nonumber \\
 \fl &=& \sum_{\left\{ \textrm{\boldmath $s$}^{(a)} \right\}} \sum_{\left\{ \textrm{\boldmath $\sigma$}^{(b)} \right\}} e^{ \beta^\prime h_0 \sum_{a=1}^m \sum_{i=1}^N s_i^{(a)} + \beta h_0 \sum_{b=1}^n \sum_{i=1}^N \sigma_i^{(b)} + \beta h_\mathrm{ext} \sum_{b=1}^n \sum_{i=1}^N \sigma_i^{(b)} s_i^{(1)}} \nonumber \\
 \fl && \qquad \times \mathbb{E}_J \left[ \frac{1}{N_G} \sum_{\left\{ g_{i_1, \cdots, i_p} \right\}} \left\{ \prod_i \delta \left( \sum_{i_1< \cdots< i_{p-1}}g_{i_1, \cdots, i_{p-1}, i} - C \right) \right\} \right. \nonumber \\
 \fl && \qquad \left. \times e^{ \sum_{i_1< \cdots< i_p} g_{i_1, \cdots, i_p} J_{i_1, \cdots, i_p} \left( \beta^\prime \sum_{a=1}^m s_{i_1}^{(a)}\cdots s_{i_p}^{(a)} + \beta \sum_{b=1}^n \sigma_{i_1}^{(b)}\cdots \sigma_{i_p}^{(b)} \right) } \right].
\end{eqnarray}
Using the relation
\begin{eqnarray}
 \fl \delta \left( \sum_{i_1< \cdots< i_{p-1}}g_{i_1, \cdots, i_{p-1}, i} - C \right) &=& \frac{1}{2\pi i}\oint dz_i z_i^{-C-1+\sum_{i_1< \cdots< i_{p-1}}g_{i_1, \cdots, i_{p-1}, i}}
\end{eqnarray}
and noticing that $g_{i_1, \cdots, i_p}$ is either $0$ or $1$, we know that
\begin{eqnarray}
 \fl \mathbb{E} \left[ Z_{n,m} \right] &=& \sum_{\left\{ \textrm{\boldmath $s$}^{(a)} \right\}} \sum_{\left\{ \textrm{\boldmath $\sigma$}^{(b)} \right\}} e^{ \beta^\prime h_0 \sum_{a=1}^m \sum_{i=1}^N s_i^{(a)} + \beta h_0 \sum_{b=1}^n \sum_{i=1}^N \sigma_i^{(b)} + \beta h_\mathrm{ext} \sum_{b=1}^n \sum_{i=1}^N \sigma_i^{(b)} s_i^{(1)}} \nonumber \\
 \fl && \qquad \times \frac{1}{N_G} \frac{1}{(2\pi i)^N} \oint \prod_i \frac{dz_i}{z_i^{C+1}} \nonumber \\
 \fl && \qquad \times \mathbb{E}_J \left[ \prod_{i_1< \cdots< i_p} \left\{ 1+z_{i_1}\cdots z_{i_p} e^{J_{i_1, \cdots, i_p} \left( \beta^\prime \sum_{a=1}^m s_{i_1}^{(a)}\cdots s_{i_p}^{(a)} + \beta \sum_{b=1}^n \sigma_{i_1}^{(b)}\cdots \sigma_{i_p}^{(b)} \right)} \right\} \right] \nonumber \\
 \fl &\sim& \sum_{\left\{ \textrm{\boldmath $s$}^{(a)} \right\}} \sum_{\left\{ \textrm{\boldmath $\sigma$}^{(b)} \right\}} e^{ \beta^\prime h_0 \sum_{a=1}^m \sum_{i=1}^N s_i^{(a)} + \beta h_0 \sum_{b=1}^n \sum_{i=1}^N \sigma_i^{(b)} + \beta h_\mathrm{ext} \sum_{b=1}^n \sum_{i=1}^N \sigma_i^{(b)} s_i^{(1)}} \nonumber \\
 \fl && \qquad \times \frac{1}{N_G} \frac{1}{(2\pi i)^N} \oint \prod_i \frac{dz_i}{z_i^{C+1}} \nonumber \\
 \fl && \qquad \times \exp \left( \sum_{\vec{s}_{(1)}, \vec{\sigma}_{(1)}} \cdots \sum_{\vec{s}_{(p)}, \vec{\sigma}_{(p)}} \frac{N^p}{p!} 2^{m+n} \cosh^m\left( \beta^\prime \left| J \right| \right) \cosh^n\left( \beta \left| J \right| \right) \right. \nonumber \\
 \fl && \qquad \times \mathbb{E}_J \left[ \prod_{a=1}^m \frac{1+\tanh\left( \beta^\prime J \right)\prod_{j=1}^p s_{(j)}^{(a)}}{2} \prod_{b=1}^n \frac{1+\tanh\left( \beta J \right)\prod_{j=1}^p \sigma_{(j)}^{(b)}}{2} \right] \nonumber \\
 \fl && \qquad \left. \times \prod_{j=1}^p \left( \frac{1}{N}\sum_{i=1}^N z_{i}\delta_{\vec{\sigma}_{(j)}, \vec{\sigma}_{i}} \delta_{\vec{s}_{(j)}, \vec{s}_{i}} \right) \right).
\end{eqnarray}
In order to obtain the equivalency, we need $N$ to be large enough.
Furthermore, when we insert the identity $1=\prod_{\vec{\sigma}, \vec{s}}\int dm\left( \vec{\sigma}, \vec{s} \right) \delta \left( m\left( \vec{\sigma}, \vec{s} \right) - \frac{1}{N}\sum_{i=1}^N z_{i}\delta_{\vec{\sigma}, \vec{\sigma}_{i}} \delta_{\vec{s}, \vec{s}_{i}} \right)$, we have
\begin{eqnarray}
 \fl \mathbb{E} \left[ Z_{n,m} \right] &=& \sum_{\left\{ \textrm{\boldmath $s$}^{(a)} \right\}} \sum_{\left\{ \textrm{\boldmath $\sigma$}^{(b)} \right\}} e^{ \beta^\prime h_0 \sum_{a=1}^m \sum_{i=1}^N s_i^{(a)} + \beta h_0 \sum_{b=1}^n \sum_{i=1}^N \sigma_i^{(b)} + \beta h_\mathrm{ext} \sum_{b=1}^n \sum_{i=1}^N \sigma_i^{(b)} s_i^{(1)}} \nonumber \\
 \fl && \qquad \times \frac{1}{N_G} \frac{1}{(2\pi i)^N} \oint \prod_i \frac{dz_i}{z_i^{C+1}} \left( \prod_{\vec{\sigma}, \vec{s}} N \int dm\left( \vec{\sigma}, \vec{s} \right) \int d\hat{m}\left( \vec{\sigma}, \vec{s} \right) \right) \nonumber \\
 \fl && \qquad \times e^{\sum_{\vec{\sigma}, \vec{s}} \hat{m}\left( \vec{\sigma}, \vec{s} \right) \left[ \sum_{i=1}^N z_{i}\delta_{\vec{\sigma}, \vec{\sigma}_{i}} \delta_{\vec{s}, \vec{s}_{i}} - N m\left( \vec{\sigma}, \vec{s} \right) \right]} \nonumber \\
 \fl && \qquad \times \exp \left( \sum_{\vec{s}_{(1)}, \vec{\sigma}_{(1)}} \cdots \sum_{\vec{s}_{(p)}, \vec{\sigma}_{(p)}} \frac{N^p}{p!} 2^{m+n} \cosh^m\left( \beta^\prime \left| J \right| \right) \cosh^n\left( \beta \left| J \right| \right) \right. \nonumber \\
 \fl && \qquad \times \left\{ \prod_{j=1}^p m\left( \vec{\sigma}_{(j)}, \vec{s}_{(j)} \right) \right\} \nonumber \\
 \fl && \qquad \left. \times \mathbb{E}_J \left[ \prod_{a=1}^m \frac{1+\tanh\left( \beta^\prime J \right)\prod_{j=1}^p s_{(j)}^{(a)}}{2} \prod_{b=1}^n \frac{1+\tanh\left( \beta J \right)\prod_{j=1}^p \sigma_{(j)}^{(b)}}{2} \right] \right).
\end{eqnarray}
Performing the integration with respect $z_i$ gives us (\ref{eq:znmcal}).

\section{Derivation of (\ref{eq:N_G_final})}
\label{ap:N_G_cal}
In this appendix, we derive (\ref{eq:N_G_final}).
Similarly to \ref{ap:znmcal}, $N_G$ is calculated as
\begin{eqnarray}
 N_G &=& \frac{1}{(2\pi i)^N} \oint \prod_i \frac{dz_i}{z_i^{C+1}} \prod_{i_1< \cdots< i_p} \left( 1 + z_{i_1}\cdots z_{i_p} \right) \nonumber \\
 &\sim& \frac{1}{(2\pi i)^N} \oint \prod_i \frac{dz_i}{z_i^{C+1}} e^{\frac{N^p}{p!} \left( \frac{1}{N}\sum_{i=1}^N z_{i} \right)^p} \nonumber \\
 &=& \frac{1}{(2\pi i)^N} \oint \prod_i \frac{dz_i}{z_i^{C+1}} \int d\nu \delta\left( \sum_iz_i -N\nu \right) e^{\frac{N^p}{p!}\nu^p} \nonumber \\
 &=& \frac{1}{(2\pi i)^N} \oint \prod_i \frac{dz_i}{z_i^{C+1}} \int d\nu \int d\hat{\nu} e^{\hat{\nu}\left( \sum_iz_i -N\nu \right)} e^{\frac{N^p}{p!}\nu^p} \nonumber \\
 &=& \int d\nu \int d\hat{\nu} \left( \frac{\hat{\nu}}{C!} \right)^N e^{-N\hat{\nu} \nu + \frac{N^p}{p!}\nu^p},
\end{eqnarray}
where we have used the fact that $N$ is sufficiently large.
By evaluating the integral with respect to $\nu$ and $\hat{\nu}$ with saddle-point values, we obtain
\begin{eqnarray}
 N_G &\sim& e^{N\left[ C\log\hat{\nu}_* -\log C! -\hat{\nu}_* \nu_* + \frac{N^{p-1}}{p!}\nu_*^p \right]},
 \label{eq:N_G_i}
\end{eqnarray}
with
\begin{eqnarray}
 \nu_*^p &=& \frac{C(p-1)!}{N^{p-1}}, \\
 \hat{\nu}_*^p &=& \frac{N^{p-1}C^{p-1}}{(p-1)!}.
\end{eqnarray}
By substituting the saddle-point values into (\ref{eq:N_G_i}), we obtain (\ref{eq:N_G_final}).

\section{Derivation of (\ref{eq:symm})}
\label{ap:ultra}
In this appendix, we derive the relation (\ref{eq:symm}).
First, we define the quantity
\begin{eqnarray}
 \hat{m}_{k,l}(s) &\equiv& \int d\hat{\rho}\left(\hat{h} \right) d\hat{\mu} \left(\hat{w}|\hat{h}, s \right) \tanh^{k}\left(\beta \hat{h} \right) \tanh^{l}\left(\beta \hat{w} \right).
\end{eqnarray}
We rewrite (\ref{eq:DEmod_P_wf}) using the Fourier transform of the delta function:
\begin{eqnarray}
 \fl \tilde{\rho} \left(\tilde{h} \right) \tilde{\mu}\left( \tilde{w}|\tilde{h}, s \right) &=& \int \frac{dy^\prime}{2\pi} e^{-iy^\prime \tilde{h}} \int \frac{dy}{2\pi} e^{-iy (\tilde{w} + \tilde{h})} \nonumber \\
 \fl && \qquad \times \left[ \sum_{n^\prime_1=0}^\infty \sum_{n^\prime_2=0}^\infty \sum_{n_1=0}^\infty \sum_{n^\prime_2=0}^\infty C\left( \frac{iy^\prime}{2\beta}, n^\prime_1 \right) C\left( -\frac{iy^\prime}{2\beta}, n^\prime_2 \right) \right. \nonumber \\
 \fl && \qquad \times C\left( \frac{iy}{2\beta}, n_1 \right) C\left( -\frac{iy}{2\beta}, n_2 \right) \nonumber \\
 \fl && \qquad \left. \times (-1)^{n^\prime_2+n_2} \hat{m}_{n^\prime_1+n^\prime_2,n_1+n_2}(s) \right]^{C-1},
 \label{eq:rhot_m}
\end{eqnarray}
in which $C\left( \alpha, n \right)$ through $(1+x)^\alpha = \sum_{n=0}^\infty C\left( \alpha, n \right)x^n$ are generalized binomial coefficients.
The quantity $\hat{m}_{k,l}(s)$ can also be rewritten using (\ref{eq:DEmod_Ph_wf}) as
\begin{eqnarray}
 \hat{m}_{k,l}(s) &=& \mathbb{I}(k+l:\mathrm{even})\left[ \sum_{m=0}^\infty \mathbb{E}_J\left[ \tanh^{k+l+2m}(\beta J)\right]q_{k+2m,l}^{p-1} \right. \nonumber \\
 && \qquad \left. -  \sum_{m=1}^\infty \mathbb{E}_J\left[ \tanh^{k+l+2m}(\beta J)\right]r_{k+2m-1,l}^{p-1} \right] \nonumber \\
 && - s \mathbb{I}(k+l:\mathrm{odd})\left[ \sum_{m=0}^\infty \mathbb{E}_J\left[ \tanh^{k+l+2m+1}(\beta J)\right]q_{k+2m+1,l}^{p-1} \right. \nonumber \\
 && \qquad \left. - \sum_{m=0}^\infty \mathbb{E}_J\left[ \tanh^{k+l+2m+1}(\beta J)\right]r_{k+2m,l}^{p-1} \right],
 \label{eq:mh_rq}
\end{eqnarray}
in which $\mathbb{I}(\cdots)$ is an indicator function that returns unity if $\cdots$ holds and vanishes otherwise. 
Equations (\ref{eq:rhot_m}) and (\ref{eq:mh_rq}) enable us to express $\tilde{\rho} \left(\tilde{h} \right) \tilde{\mu}\left( \tilde{w}|\tilde{h}, s \right)$ using $r_{k,l}$ and $q_{k,l+1}$.
Using $\hat{m}_{k,l}(s)$, we derive
\begin{eqnarray}
 \fl && \tilde{\rho} \left(\tilde{h} \right) \tilde{\mu}\left( \tilde{w}|\tilde{h}, 1 \right) e^{-2\beta \tilde{w}} \nonumber \\
 \fl &=& \int \frac{dy^\prime}{2\pi} e^{-iy^\prime \tilde{h}} \int \frac{dy}{2\pi} e^{-iy (\tilde{w} + \tilde{h})} \left[ \sum_{n^\prime_1=0}^\infty \sum_{n^\prime_2=0}^\infty \sum_{n_1=0}^\infty \sum_{n^\prime_2=0}^\infty C\left( \frac{iy^\prime}{2\beta}, n^\prime_1 \right) C\left( -\frac{iy^\prime}{2\beta}, n^\prime_2 \right) \right. \nonumber \\
 \fl && \qquad \times C\left( \frac{iy}{2\beta}, n_1 \right) C\left( -\frac{iy}{2\beta}, n_2 \right) (-1)^{n^\prime_2+n_2} \nonumber \\
 \fl && \qquad \times \left\{ \hat{m}_{n^\prime_1+n^\prime_2,n_1+n_2}(1) + 2\sum_{l^\prime=0}^\infty \hat{m}_{n^\prime_1+n^\prime_2+2l^\prime+2,n_1+n_2}(1) \right. \nonumber \\
 \fl && \qquad + 2\sum_{l^\prime=0}^\infty \hat{m}_{n^\prime_1+n^\prime_2+2l^\prime+1,n_1+n_2}(1) + 2\sum_{l=0}^\infty \hat{m}_{n^\prime_1+n^\prime_2,n_1+n_2+2l+2}(1) \nonumber \\
 \fl && \qquad + 4\sum_{l=0}^\infty \sum_{l^\prime=0}^\infty \hat{m}_{n^\prime_1+n^\prime_2+2l^\prime+2,n_1+n_2+2l+2}(1) \nonumber \\
 \fl && \qquad + 4\sum_{l=0}^\infty \sum_{l^\prime=0}^\infty \hat{m}_{n^\prime_1+n^\prime_2+2l^\prime+1,n_1+n_2+2l+2}(1) - 2\sum_{l=0}^\infty \hat{m}_{n^\prime_1+n^\prime_2,n_1+n_2+2l+1}(1) \nonumber \\
 \fl && \qquad - 4\sum_{l=0}^\infty \sum_{l^\prime=0}^\infty \hat{m}_{n^\prime_1+n^\prime_2+2l^\prime+2,n_1+n_2+2l+1}(1) \nonumber \\
 \fl && \qquad \left. \left. - 4\sum_{l=0}^\infty \sum_{l^\prime=0}^\infty \hat{m}_{n^\prime_1+n^\prime_2+2l^\prime+1,n_1+n_2+2l+1}(1) \right\} \right]^{C-1}.
 \label{eq:ultra_cal}
\end{eqnarray}
Now we assume the relation (\ref{eq:ultra}).
Then we can apply (\ref{eq:mh_rq}) in order to obtain
\begin{eqnarray}
 \fl && \sum_{l^\prime=0}^\infty \left[ \hat{m}_{n^\prime_1+n^\prime_2+2l^\prime+2,n_1+n_2}(1) + \hat{m}_{n^\prime_1+n^\prime_2+2l^\prime+1,n_1+n_2}(1) \right] \nonumber \\
 \fl &=& \mathbb{I}(n^\prime_1+n^\prime_2+n_1+n_2:\mathrm{odd})\sum_{l^\prime=0}^\infty \mathbb{E}_J\left[ \tanh^{n^\prime_1+n^\prime_2+2l^\prime+1+n_1+n_2}(\beta J)\right] \nonumber \\
 \fl && \qquad \times q_{n^\prime_1+n^\prime_2+2l^\prime+1,n_1+n_2}^{p-1} \nonumber \\
 \fl && + \mathbb{I}(n^\prime_1+n^\prime_2+n_1+n_2:\mathrm{even})\sum_{l^\prime=0}^\infty \mathbb{E}_J\left[ \tanh^{n^\prime_1+n^\prime_2+2l^\prime+2+n_1+n_2}(\beta J)\right] \nonumber \\
 \fl && \qquad \times q_{n^\prime_1+n^\prime_2+2l^\prime+1,n_1+n_2+1}^{p-1}.
\end{eqnarray}
Similarly, we also have
\begin{eqnarray}
 \fl && \sum_{l=0}^\infty \left[ \hat{m}_{n^\prime_1+n^\prime_2,n_1+n_2+2l+2}(1) - \hat{m}_{n^\prime_1+n^\prime_2,n_1+n_2+2l+1}(1) \right] \nonumber \\
 \fl &=& -\mathbb{I}(n^\prime_1+n^\prime_2+n_1+n_2:\mathrm{odd})\sum_{l=0}^\infty \mathbb{E}_J\left[ \tanh^{n^\prime_1+n^\prime_2+2l+1+n_1+n_2}(\beta J)\right] \nonumber \\
 \fl && \qquad \times q_{n^\prime_1+n^\prime_2+2l,n_1+n_2+1}^{p-1} \nonumber \\
 \fl && + \mathbb{I}(n^\prime_1+n^\prime_2+n_1+n_2:\mathrm{even})\sum_{l=0}^\infty \mathbb{E}_J\left[ \tanh^{n^\prime_1+n^\prime_2+2l+2+n_1+n_2}(\beta J)\right] \nonumber \\
 \fl && \qquad \times q_{n^\prime_1+n^\prime_2+2l+1,n_1+n_2+1}^{p-1}.
\end{eqnarray}
Furthermore, we know
\begin{eqnarray}
 \fl && \sum_{l=0}^\infty \sum_{l^\prime=0}^\infty \left[ \hat{m}_{n^\prime_1+n^\prime_2+2l^\prime+2,n_1+n_2+2l+2}(1) + \hat{m}_{n^\prime_1+n^\prime_2+2l^\prime+1,n_1+n_2+2l+2}(1) \right. \nonumber \\
 \fl && \qquad \left. - \hat{m}_{n^\prime_1+n^\prime_2+2l^\prime+2,n_1+n_2+2l+1}(1) - \hat{m}_{n^\prime_1+n^\prime_2+2l^\prime+1,n_1+n_2+2l+1}(1) \right] \nonumber \\
 \fl &=& - \mathbb{I}(n^\prime_1+n^\prime_2+n_1+n_2:\mathrm{even}) \sum_{l^\prime=0}^\infty \mathbb{E}_J\left[ \tanh^{n^\prime_1+n^\prime_2+2l^\prime+1+n_1+n_2+1}(\beta J)\right] \nonumber \\
 \fl && \qquad \times q_{n^\prime_1+n^\prime_2+2l^\prime+1,n_1+n_2+1}^{p-1}.
\end{eqnarray}
Combining these results gives us
\begin{eqnarray}
 \fl && \hat{m}_{n^\prime_1+n^\prime_2,n_1+n_2}(1) + 2\sum_{l^\prime=0}^\infty \hat{m}_{n^\prime_1+n^\prime_2+2l^\prime+2,n_1+n_2}(1) + 2\sum_{l^\prime=0}^\infty \hat{m}_{n^\prime_1+n^\prime_2+2l^\prime+1,n_1+n_2}(1) \nonumber \\
 \fl && \qquad + 2\sum_{l=0}^\infty \hat{m}_{n^\prime_1+n^\prime_2,n_1+n_2+2l+2}(1) + 4\sum_{l=0}^\infty \sum_{l^\prime=0}^\infty \hat{m}_{n^\prime_1+n^\prime_2+2l^\prime+2,n_1+n_2+2l+2}(1) \nonumber \\
 \fl && \qquad + 4\sum_{l=0}^\infty \sum_{l^\prime=0}^\infty \hat{m}_{n^\prime_1+n^\prime_2+2l^\prime+1,n_1+n_2+2l+2}(1) - 2\sum_{l=0}^\infty \hat{m}_{n^\prime_1+n^\prime_2,n_1+n_2+2l+1}(1) \nonumber \\
 \fl && \qquad - 4\sum_{l=0}^\infty \sum_{l^\prime=0}^\infty \hat{m}_{n^\prime_1+n^\prime_2+2l^\prime+2,n_1+n_2+2l+1}(1) - 4\sum_{l=0}^\infty \sum_{l^\prime=0}^\infty \hat{m}_{n^\prime_1+n^\prime_2+2l^\prime+1,n_1+n_2+2l+1}(1) \nonumber \\
 \fl &=& \hat{m}_{n^\prime_1+n^\prime_2,n_1+n_2}(-1).
\end{eqnarray}
By substituting this into (\ref{eq:ultra_cal}) and comparing it to (\ref{eq:rhot_m}) with $s=-1$, we obtain (\ref{eq:symm}).

\end{document}